\documentclass[twoside,twocolumn,english,amsmath,amssymb,superscriptaddress,nofootinbib]{revtex4-1}
\usepackage[T1]{fontenc}
\usepackage[latin9]{inputenc}
\usepackage{babel}
\usepackage{amsmath}
\usepackage{amssymb}
\usepackage{graphicx}
\usepackage{esint}
\usepackage[unicode=true,pdfusetitle,
 bookmarks=true,bookmarksnumbered=true,bookmarksopen=false,
 breaklinks=false,pdfborder={0 0 1},backref=false,colorlinks=false]
 {hyperref}
 
\makeatletter
\usepackage{babel}
\usepackage{newtxtext}

\makeatother

\begin{document}

\title{Monte-Carlo Simulations of Superradiant Lasing}

\author{Yuan Zhang}
\email{yzhang@phys.au.dk}

\selectlanguage{english}%

\address{Department of Physics and Astronomy, Aarhus University, Ny Munkegade
120, DK-8000 Aarhus C, Denmark}

\author{YuXiang Zhang}
\email{iyxz@phys.au.dk}

\author{Klaus M{\o}lmer}
\email{moelmer@phys.au.dk}

\selectlanguage{english}%

\address{Department of Physics and Astronomy, Aarhus University, Ny Munkegade
120, DK-8000 Aarhus C, Denmark}
\begin{abstract}
We present simulations of the superradiant dynamics of ensembles of atoms in the presence of collective and individual atomic decay processes.  We unravel the density matrix with Monte-Carlo wave-functions and identify the quantum jumps in a reduced Dicke state basis, which reflects the permutation symmetry of the identical atoms. While the number of density matrix elements in the Dicke representation increases polynomially with atom number, the quantum jump dynamics populates only a single Dicke state at the time and thus efficient simulations can be carried out for tens of thousands of atoms. The calculated superradiance pulses from initially excited atoms agree quantitatively with recent experimental results with strontium atoms but rapid atom loss in these experiments does not permit steady-state superradiance. By introducing an incident flux of new atoms, the system can maintain a large average atom number, and our theoretical calculations predict lasing with millihertz linewidth despite rapid atom number fluctuations.
\end{abstract}
\maketitle

\paragraph*{Introduction}
Superradiance is caused by the collective interaction of atoms with a common radiation field and has been the subject of continuous interest since the early proposal by Dicke \cite{RHDicke1}, (see also \cite{AVandreev,BarryMGarraway}). Recent experiments with tens of thousands of atoms in optical cavities \cite{MatthewAN,MANorcia1,JGBohnet} thus explore the possibility of achieving superradiant lasing in a bad cavity. Opposite to the Schawlow-Townes limit of a normal laser, it has a linewidth set by the rate of energy transfer from single atoms to the cavity mode and can in principle reach millihertz level \cite{DMeiser}. However, so far, such performance has been hindered by fast atom loss depleting the atomic ensemble before the coherence is established \cite{MatthewAN}. To solve this problem, we might feed new atoms to the system to stabilize the atom number. This will invariably introduce noise in the system but the coherence shared among the atoms might still be preserved over the time scale where all atoms have been replaced.

\begin{figure}
\begin{centering}
\includegraphics[scale=1.0]{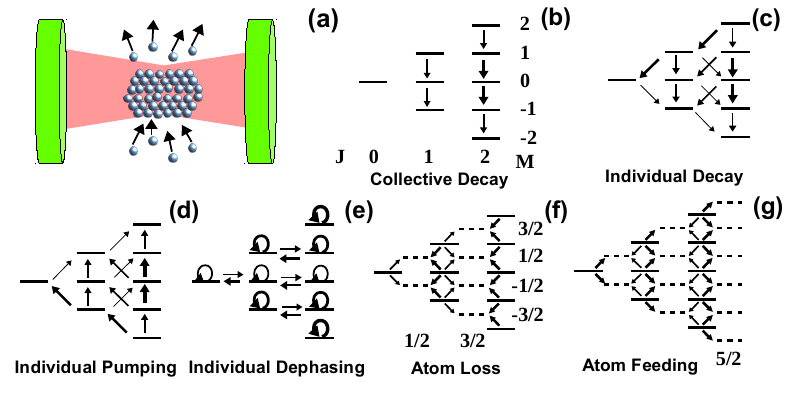}
\par\end{centering}
\caption{\label{fig:system-process}System and processes. Panel (a) shows two-level atoms located within an optical cavity in the presence of the loss of atoms from the cavity mode volume and the feeding of new atoms into the volume. Panels (b-g) show Dicke state diagrams for four atoms illustrating quantum jumps due to collective decay (b);  decay, pumping, and dephasing of individual atoms (c-e); atom loss and feeding (f,g). The thickness of the arrows indicates the probabilities of the different jumps as discussed in the text.}
\end{figure}

In this Letter, we perform numerical studies of the system depicted in Fig. \ref{fig:system-process} (a), where atoms interact collectively with a lossy cavity mode while being subject to individual decay, dephasing and excitation (pumping) due to the interaction with their local environment, as well as to loss and feeding. Theoretically, this system can be studied with either the laser master equation or, by elimination of the bad cavity, an atomic superradiance master equation \cite{RBonifacio}. However, since the number of density matrix elements scales exponentially with the atom number, the calculations in the atomic product states basis are normally restricted to only tens of atoms \cite{DMeiser1}. Assuming identical interactions and permutation symmetry among all the atoms, calculations can be carried out in a Dicke states basis \cite{BAChase,BAChase1,FDamanet} or collective numbers basis \cite{MRichter,MGegg,YZhang} (exploiting  SU(4) group theory \cite{MXu}), where the complexity scales only cubically with the atom number and thus permits simulations for hundreds of atoms. In this Letter we employ the Monte Carlo wave-function method (MCWF) to unravel the master equation in the Dicke states basis. This, on the one hand, provides novel insights into the excitation dynamics of the system and, on the other hand, permits simulations for tens of thousands of atoms as encountered in experiments.

The MCWF method \cite{KMoeler,MBPlenio} utilizes ensembles of wave-functions
instead of a density matrix to represent a quantum system, and applies random quantum jumps to describe the effect of dissipation. The jumps can also describe back-action associated with detection, and thus form an essential component of so-called quantum trajectories \cite{HJCarmichael} with applications in quantum measurement and control \cite{HMWiseman}. When a system is allowed to explore all states of a high dimensional Hilbert space, the conditional wave-functions are lower dimensional objects than density matrices. This fact renders the MCWF method an efficient numerical tool for complex system simulations. Furthermore, the symmetries mentioned above restrict the dynamics to sub-spaces and thus can potentially speed up the wave-function calculations.

The theoretical treatment of $N$ atoms that start in the same pure state and experience only collective decay, is restricted to the sub-space of fully symmetric Dicke states $|J,M\rangle$ with collective spin quantum numbers $J=N/2$ and $M=-J, ... J$ representing the total number of excited atoms $J+M$. The individual decay, dephasing and incoherent excitation as well as the atom loss and feeding seemingly break the symmetry among the atoms. However, if all atoms undergo these processes with the same rates, the density matrix retains its symmetry under permutation of the atoms, but will explore the Dicke states $|J,M\rangle$ with $J=N/2, N/2-1, ... 1/2$ or $0$, and $M=-J, ... J$ \cite{NShammah}. In the following, we review briefly the master equation for the density matrix in the Dicke states basis and then demonstrate its unraveling and effective simulation with the MCWF method.

\paragraph*{Superradiance Master Equation and Monte Carlo Wave-Function Simulations}

The master equation for the density matrix of $N$ two-level atoms with identical transition frequency $\omega_{a}$ can be written as
\begin{align}
\frac{\partial}{\partial t}\rho & =-\frac{i}{\hbar}\left[H_{a},\rho\right]+\mathcal{D}^{c}\left[\rho\right]+\mathcal{D}^{l}\left[\rho\right],\label{eq:SME}
\end{align}
where the Hamiltonian $H_{a}=\hbar\left(\omega_{a}/2\right)\sigma^{z}$
is characterized by the collective Pauli operator $\sigma^{z}=\sum_{k=1}^{N}\sigma_{k}^{z}$. The cavity mode with frequency $\omega_c$ and loss rate $\gamma_c$ provides a channel for the collective decay, and in the bad cavity limit the elimination of the cavity field operators yields  \cite{RBonifacio}  $\mathcal{D}^{c}\left[\rho\right]=-i\left[\omega_{s}\sigma^{+}\sigma^{-},\rho\right]
-\Gamma_{c}\mathcal{D}\left[\sigma^{-}\right]\rho$ with collective raising (lowering) operator $\sigma^+ = \sum_k \sigma^{+}_k$  ($\sigma^- = \sum_k \sigma^{-}_k$) and  superoperator $\mathcal{D}\left[o\right]\rho=\left\{ o^{+}o,\rho\right\} /2-o\rho o^{+}$. Here,  $\omega_{s}=g^{2}\chi\left[\chi^{2}+\gamma_c^{2}/4\right]^{-1}$
 and $\Gamma_{c}=\left(g^{2}\gamma_c/2\right)\left[\chi^{2}+\gamma_c^{2}/4\right]^{-1}$ are the collective Lamb shift  and the collective decay rate, respectively, and are determined by the detuning $\chi=\omega_c-\omega_a$, the cavity loss $\gamma_c$ and the atom-cavity mode coupling $g$. The last term in Eq. (\ref{eq:SME})  represents the dissipation of individual atoms by three Lindblad terms $\mathcal{D}^{l}\left[\rho\right]=- \gamma_l \sum_{k}\mathcal{D}\left[\sigma_{k}^{-}\right]\rho- \kappa_l \sum_{k}\mathcal{D}\left[\sigma_{k}^{+}\right]\rho- d_l \sum_{k}\mathcal{D}\left[\sigma_{k}^{z}\right]\rho$ with a decay rate $\gamma_l$, a pumping rate $\kappa_l$
and a dephasing rate $d_l$. To realize the incoherent pumping with $\kappa_l$ and atomic population inversion, the atoms may be excited from the lower level via a higher excited level from which they decay rapidly to the upper level of our effective two-level systems \cite{DMeiser,DMeiser1}.
Further below, we shall generalize our analysis to the case with varying atom number.

Dicke states $\left|JM\right\rangle $ are common eigenstates of the collective operators
$j^{2}=\sum_{l=x,y,z}\left(j^{l}\right)^{2}$
and $j^{z}$  \cite{RHDicke1} with eigenvalues $J(J+1)$ and $M$, where $j^{x}=(\sigma^{+} +\sigma^{-})/2$, $j^{y}=i( \sigma^{-}-\sigma^{+})/2$  and $j^{z}=\sigma^{z}/2$.
For $J < N/2$, the states with same $J$ and $M$ are degenerate, and may be labeled by an additional quantum number $\alpha=1, ...   , d_{N}^{J}$, which specifies $d_{N}^{J}=N!\left(2J+1\right)/\left[\left(N/2-J\right)!\left(N/2+J+1\right)!\right]$ different symmetric linear combinations of atomic product states \cite{LMandel}. Due to the identical dissipation of all the atoms, it was shown in  \cite{BAChase,BAChase1} that while the dissipative processes of a single atom break the symmetry of the many body state, the sum of these processes over all the atoms in Eq. (\ref{eq:SME}) preserves the symmetry and populates states with different $\alpha$ evenly. These processes do not build coherence between states of different $\alpha$ or $J$, and the equality of density matrix elements $\rho_{JM\alpha,JM'\alpha}$ for different $\alpha$ motivates us to introduce a single term $\rho^J_{MM'}= (1/d_N^J) \sum_\alpha \rho_{JM\alpha,JM'\alpha}$ to represent all of these elements. The resulting effective density matrix is normalized as Tr$(\rho) \equiv \sum_{J,M} d_{N}^{J} \rho^J_{MM} = 1$. Notice that the degeneracy factor appears also in the evaluation of physical observables.

The equations for all the specified density matrix elements are derived in \cite{BAChase,BAChase1}. Now we proceed to unravel them with the MCWF method. To this end we introduce an ensemble of wave-functions
\begin{equation}
\left|\psi_i \left(t\right)\right\rangle = \sum_{J,M}C_{JM}^i \left(t\right) \left|JM\right\rangle. \label{eq:Wave-Function}
\end{equation}
Here, the effective states $\left|JM\right\rangle$ are not real states, but symbolic constructions, which disregard the degeneracy of the subspace with same $J$ and $M$, but allow sampling of the density matrix element $\rho^J_{MM'}$ as $\frac{1}{n} \sum_{i=1}^n  C_{JM}^{i} C^{i*}_{JM'}$. The normalization of the wave-functions is $\sum_{J,M} d_{N}^{J} |C_{JM}^{i}|^2 = 1$. The absence of coherence between different  $J$ allows us to consider only the matrix elements and thus amplitudes of the wave-functions  for each separate value of $J$. 

In the Appendices, we detail the evaluation of the Monte Carlo wave-functions. Here, we  emphasize the main procedures and highlight the related quantum jumps.  The wave-functions are represented as state vectors in the Dicke states space and are propagated with a non-Hermitian Hamiltonian including the atomic Hamiltonian and the anti-commutator of the dissipative superoperators. The $C_{JM}^i$ amplitudes follow separate equations including a reduction of norm associated with different physical processes that we simulate by discrete quantum jumps.  The collective decay applies the jump operator $\sigma^-$ on the state-vectors, which leads to the jumps to the Dicke states with the same $J$ but reduced $M$, cf. Fig. \ref{fig:system-process} (b),  with a probability proportional to $|A_-^{JM}|^2$  [with $A_-^{JM}=\sqrt{(J+M)(J-M+1)}$ ]. 

For the individual decay, we use the Clebsch-Gordan (CG) expansion \cite{BAChase,BAChase1} to write the Dicke states of $N$ atoms as the product of those of $N-1$ atoms and a single atom, and then apply the operator $\sigma_k^-$ on the single atom state. Finally, we use the inverse CG expansion to convert the state back to the Dicke states basis of all $N$ atoms. This introduces jumps to Dicke states with $M$ reduced by unity and $J$ changed by $s=0,+1,-1$. The branching ratio of the jumps are given in the Table \ref{jumps} in the Appendices and favors jumps to states with reduced values of $J$, cf. Fig. \ref{fig:system-process}(c). We apply the same treatment to the individual pumping and dephasing by applying the operators $\sigma_k^+$ and $\sigma_k^z$ to the Dicke states, yielding jumps to states with quantum numbers $J+s$  and $M+1$ and $M$, cf. Fig. \ref{fig:system-process} (d,e) and the Table \ref{jumps} in the Appendices.

We now consider the quantum jump description of loss of atoms and feeding of new atoms into the system. For that purpose, we assume the atom loss and feeding rates, $\gamma_{loss}$ and $\kappa_{feed}$, are independent of the atomic internal state. The average atom number follows the rate equation  ${\partial N}/{\partial t} = -\gamma_{loss}N + \kappa_{feed}$, which yields a steady state Poisson distribution of atom number with mean  $N_s = \kappa_{feed}/\gamma_{loss}$. To simulate the loss of an atom we use the CG expansion to decouple the Dicke states of $N$ atoms as a product of states of  $N-1$ atoms and of a single atom, and then we discard the state of the single atom by a partial trace, simulated by quantum jumps from Dicke states of $N$ atoms to those of $N-1$ atoms with probabilities proportional to squared CG coefficients, cf. Fig. \ref{fig:system-process} (f) and the Appendices. To incorporate a new atom, we use the inverse CG expansion and simulate a quantum jump from the Dicke state of $N$ atoms to one of $N+1$ atoms, cf. Fig. \ref{fig:system-process} (g) and the Appendices. After the implementation of any of the above quantum jumps, the state vectors are re-normalized, and the wave-function propagation and further random jumps proceed.

The number of the effective Dicke states of $N$ atoms is $\left(N+3\right)\left(N+1\right)/4$
for odd $N$ and $\left(N+2\right)^{2}/4$ for even $N$. However, if the system has no initial coherence between states of different $J$, such a coherence will never appear since all the jumps happen either within the same ladder or from one to a neighboring ladder, cf. Fig.\ref{fig:system-process} (b-g). Thus, our wave-functions can at any time be expanded on only a single ladder of states, restricting the dimension to $2J+1 \leq N+1$ elements. If the system is initially prepared in a single Dicke state, e.g. the ground or fully-excited state, all coherences between states with different $M$ vanish for all times, and the system evolution can be effectively simulated as an incoherent jump process between states $|JM\rangle$. In this case, the dynamics of a stochastic wave-function can be visualized as a trajectory in a $(N,J,M)$ coordinate system, and we only need to evaluate the different jump probabilities to simulate the random evolution of $N(t),J(t)$, and $M(t)$. By simulation of multiple such trajectories we can evaluate ensemble average of physical observables, such as the population of the Dicke states $P^i_{JM} = |C^i_{JM}(t)|^2$, and the radiation through the cavity mode $ \Gamma_c \sum_{JM} d_N^J (A_-^{JM})^2 P^i_{JM}$ (defined with the collective operators $\sigma^+\sigma^-$), and the radiation into the free space $\gamma_l \sum_{JM} d_N^J (N/2+M) P^i_{JM}$ (defined with the operators $\sum_k \sigma^+_k \sigma^-_k$). 

\paragraph*{Superradiant Dynamics of a Fixed Number of Atoms}

\begin{figure}
\begin{centering}
\includegraphics[scale=0.25]{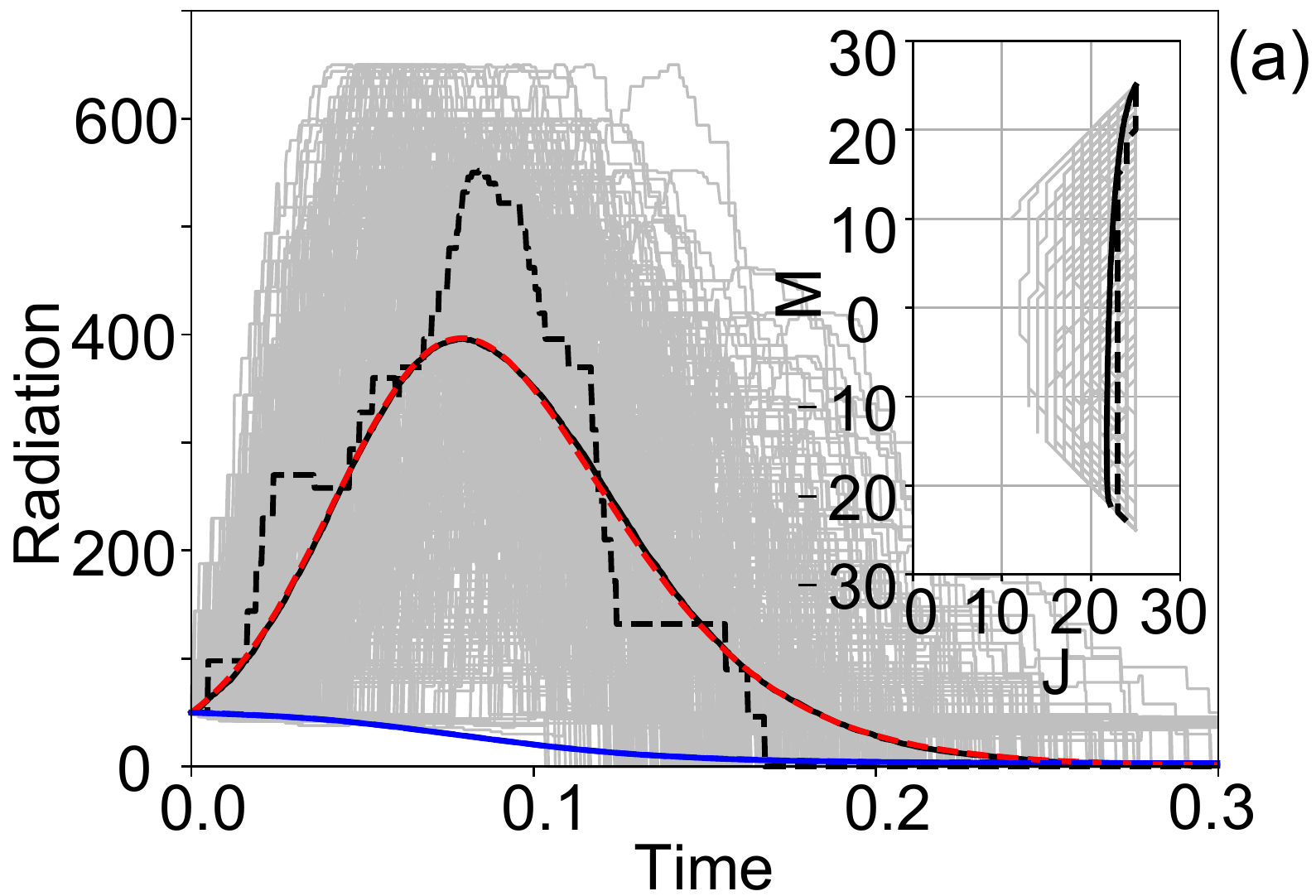} \includegraphics[scale=0.25]{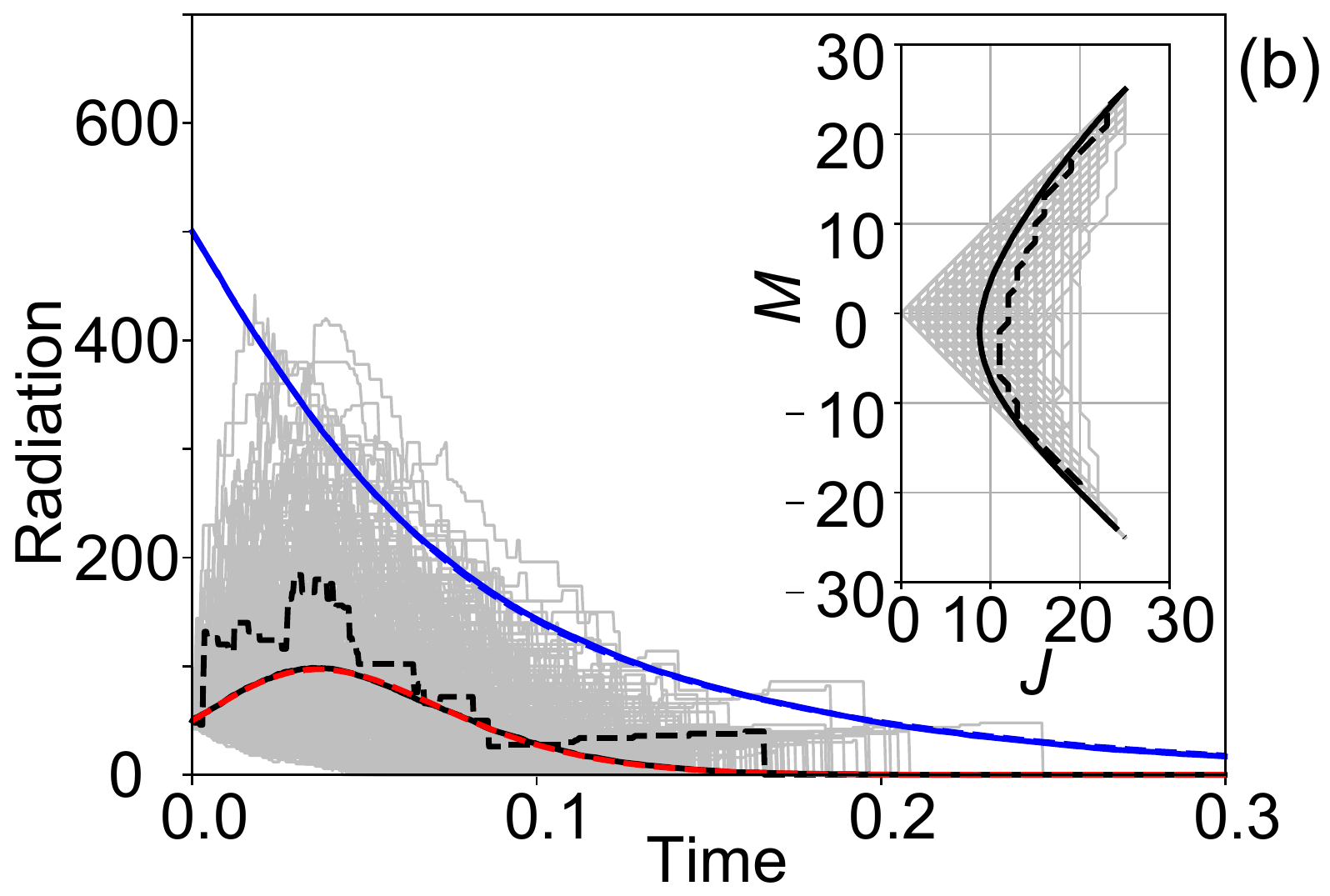}
\par\end{centering}
\includegraphics[scale=0.25]{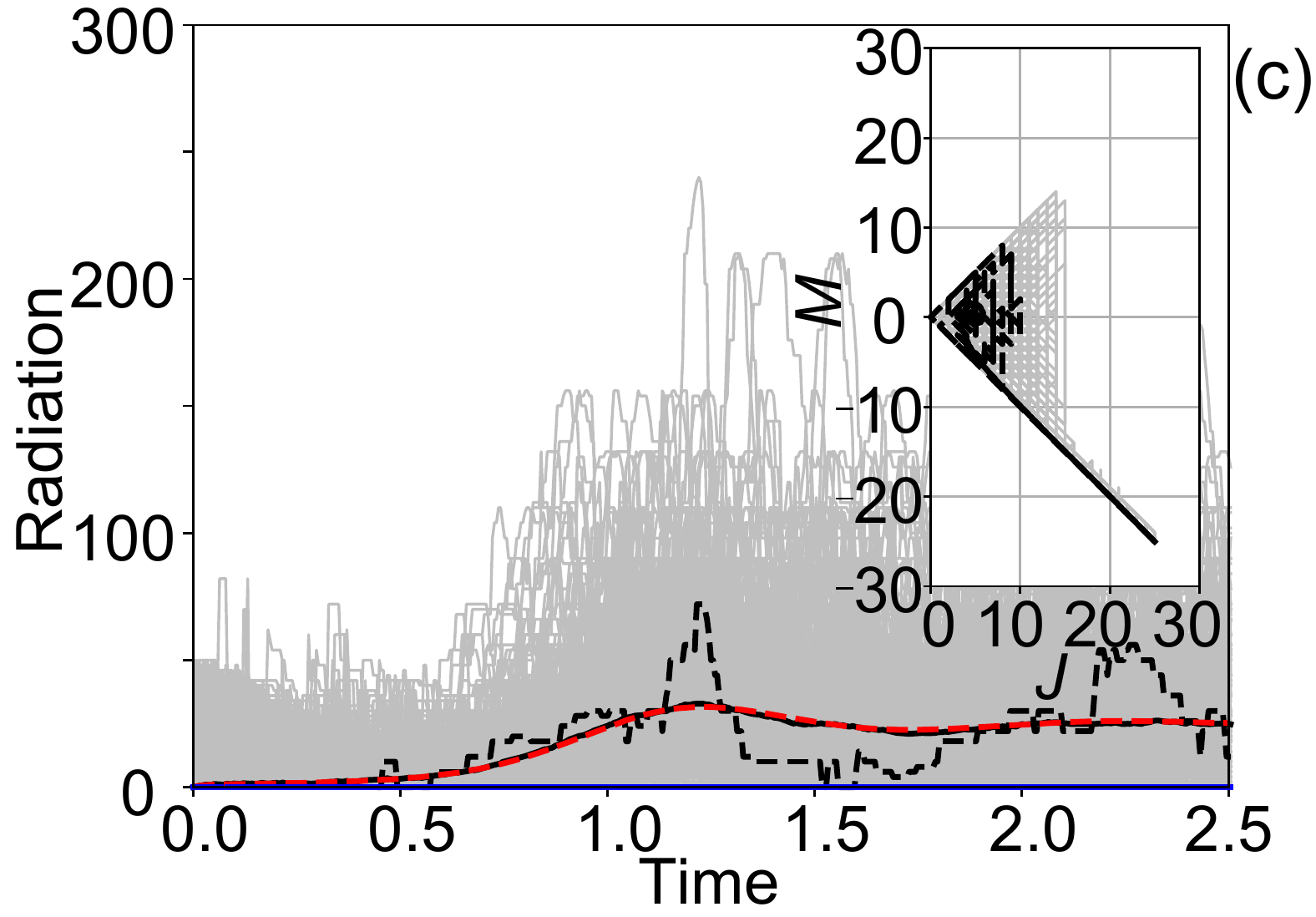} \includegraphics[scale=0.25]{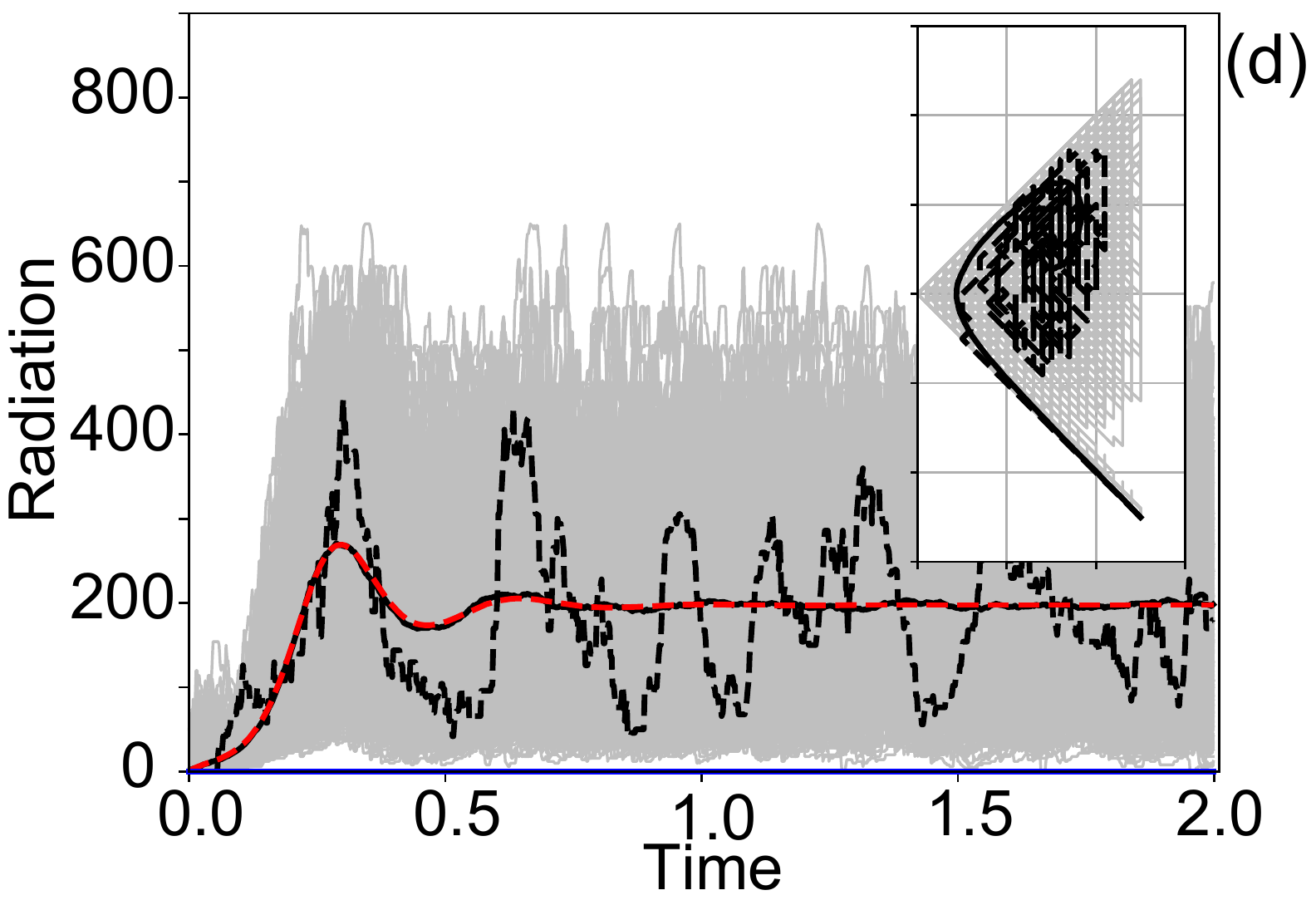}
\caption{\label{fig:superradiance-dynamics} Dynamics and radiation of fifty atoms
initially excited (a,b) and initially unexcited but continuously pumped (c,d). The main panels show the radiation intensity (in unit of the collective decay rate $\Gamma_c$) versus the dimensionless time $t \Gamma_c$ while their insets show ensembles of quantum jump trajectories among the $|JM\rangle$ Dicke states. The gray areas are explored by $512$ simulated trajectories, while single sample trajectories and average trajectories are indicated by the dashed and solid black curves, respectively. The red dashed curves are computed with the master equation and match perfectly with the black solid curves. The blue solid curves show the averaged  radiation into the free-space.  $\gamma_l/\Gamma_c = 1,10$ for the panels (a,b) while $\kappa_l/\Gamma_c = 1,10$ for the panels (c,d). }
\end{figure}

We first illustrate our simulations with a small system of fifty atoms (in the absence of atom loss and feeding), which are initially all excited, i.e. we occupy the Dicke state $\left|J=25,M=25\right\rangle$, cf. Fig. \ref{fig:superradiance-dynamics} (a,b), and initially all in the ground state, i.e. we occupy the Dicke state $\left|J=25,M=-25\right\rangle$, but pumped incoherently, cf. Fig. \ref{fig:superradiance-dynamics} (c,d). The main panels show the time evolution of the emitted radiation intensity, while the insets show the quantum jump trajectories among Dicke states.

To study the influence of individual decay with rate $\gamma_l$, we set $\gamma_l=\Gamma_c$ in Fig. \ref{fig:superradiance-dynamics} (a) and $\gamma_l=10 \Gamma_c$ in Fig. \ref{fig:superradiance-dynamics} (b). For weak individual decay, we observe that the system remains close to the fully symmetric Dicke states with $J=N/2$, while for larger individual decay, it departs from these states and explores states with lower values of $J$. In all cases, all atoms eventually end up in the ground state $\left|J=25,M=-25\right\rangle$ in the lower right corner in both insets. This dynamics is also reflected in the radiative emission by the system. Figure \ref{fig:superradiance-dynamics} (a) shows a minor radiation into the free-space (the blue curve), and a dominant radiation through the cavity mode, i.e. superradiance. The different gray curves reflect shot-to-shot variations in the superradiance from the system, which are rather large due to the small number of atoms in our simulations. The average superradiance (the black solid curve) agrees well with the master equation results (the red dashed curve), verifying our MCWF method. Figure \ref{fig:superradiance-dynamics} (b) shows that for strong individual decay, the superradiance is suppressed and weaker than the radiation into the free-space (the blue curve), which follows the usual exponential decay law. 

To study the effect of individual pumping with rate $\kappa_l$, we set $\kappa_l = \Gamma_c$ in Fig. \ref{fig:superradiance-dynamics}(c), and $\kappa_l =10\Gamma_c$ in Fig. \ref{fig:superradiance-dynamics}(d). For weak pumping the trajectories start from the lower right corner and propagate along the lower boundary of the Dicke ladders, and finally wander around the left corner with small values of $J,M$, while the radiation through the cavity mode fluctuates around a steady-state mean value $25\Gamma_{c}$. For strong pumping the trajectories explore more states and end up in a region around  $J=15$ and $M=8$.  As a consequence of the higher symmetry of the states and their higher degree of excitation, the collectively enhanced emission probabilities are higher and both the fluctuations and the averaged radiation increase. For even stronger pumping, we find that the trajectories propagate to the upper right corner of the inset Dicke state diagram, and the radiation through the cavity mode is reduced. In all cases, the averaged radiation trajectories are in perfect agreement with the exact results calculated with the master equation. The reduced radiation at strong pumping can be ascribed to the effect of the pump-induced noise \cite{DMeiser}, and may also be understood as a suppression of the superradiance transition matrix elements $\propto A^{JM}_-= \sqrt{(J+M)(J-M+1)}$ when the atoms do not occupy both the ground and excited states.

\begin{figure}
\begin{centering}
\includegraphics[scale=0.45]{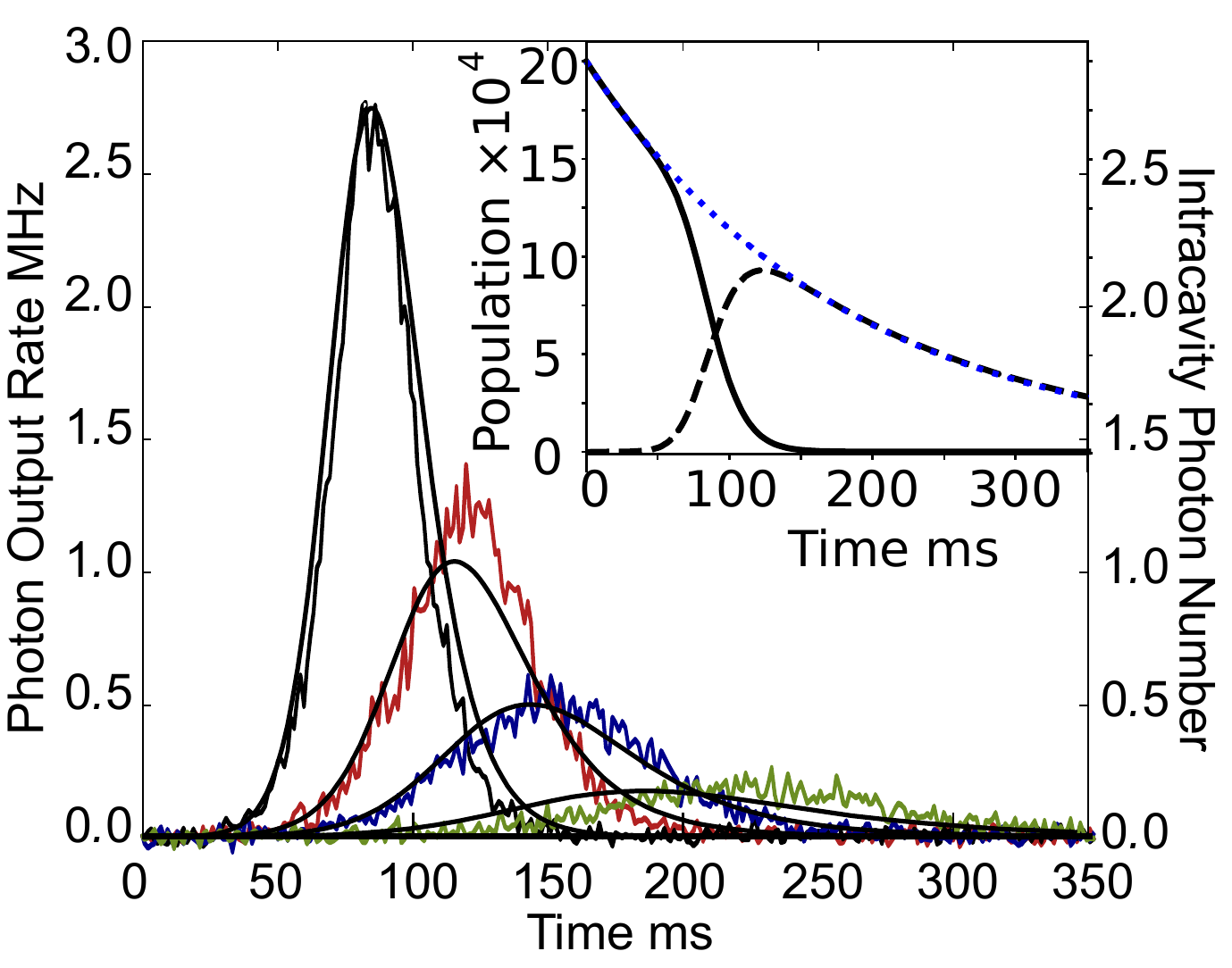}
\par\end{centering}
\caption{\label{fig:superradiance-pulse} Superradiance pulses of $N(0)=1.0,1.25,1.5,2.0 \times10^{5}$ strontium atoms (from right to left smooth curve), which decay with a loss rate $\gamma_{loss} = 5.58$ Hz (the noisy curves are experimental results from \cite{MatthewAN}). The inset shows the population of the upper (black solid curve) and lower atomic level (black dashed curve) as well as the total atom number (blue dotted curve) for the system with $N(0)=2\times 10^5$ atoms. $\gamma_l = 6.6$ mHz, $\Gamma_{c} = d_l =0.91$ mHz. }
\end{figure}

\begin{figure}
\begin{centering}
\includegraphics[scale=0.45]{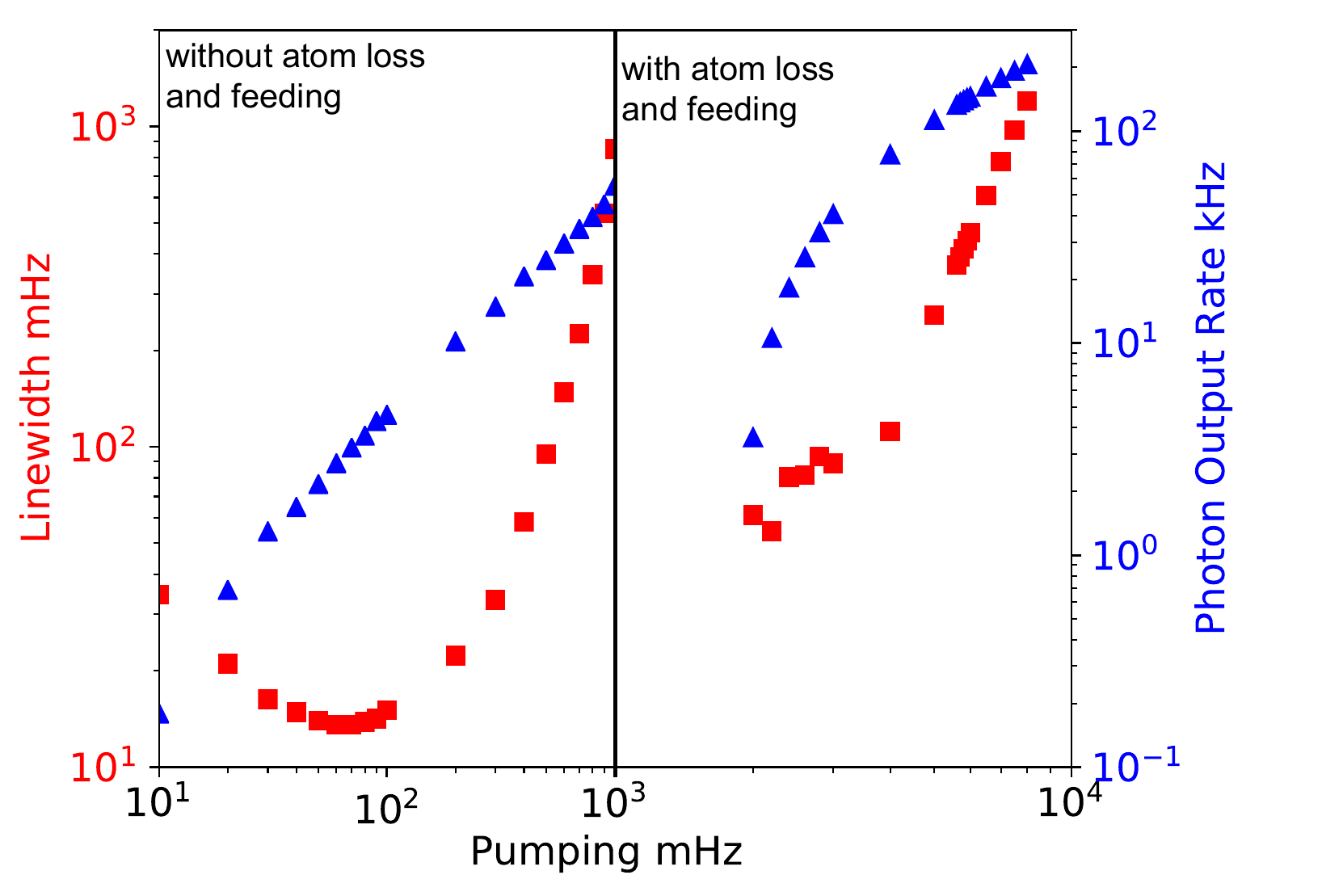}
\par\end{centering}
\caption{\label{fig:steady-state-superradiance} Integrated intensity (blue triangles, right axis) and linewidth (red squares, left axis) of the steady-state spectrum from $N_s=10^{5}$  strontium atoms versus pumping rate $\kappa_l$ without (left) and with (right) atom loss and feeding (of atoms in the lower state). $\kappa_{feed}= \gamma_{loss} N_s$ and other parameters are the same as in Fig. \ref{fig:steady-state-superradiance}.}
\end{figure}

\paragraph*{Superradiant Lasing}
After having verified our method with a toy model calculation for tens of atoms, we now turn to the simulation of much larger ensembles, cf. Fig. \ref{fig:superradiance-pulse} and \ref{fig:steady-state-superradiance}, as studied by Norcia et. al. \cite{MatthewAN}. In their
experiment, more than $10^{5}$ strontium atoms
are trapped in an optical lattice inside an optical cavity and coupled
to a cavity mode through their ultra-narrow $S_{0}^{1}-P_{0}^{3}$ optical
clock transition.

For this problem, the many independent trajectories explore ranges of $J$ and $M$ according to the same diffusion-like process as observed in Fig. \ref{fig:superradiance-dynamics}. However, with $N > 10^5$ we have verified that the relative fluctuations among trajectories remain very small, and hence we can simulate the system and obtain the information of interest with only few hundred quantum trajectories. In Fig. \ref{fig:superradiance-pulse} we show the superradiance pulses for different numbers of initial atoms $N(0)$ (from $1\times10^{5}$ to $2\times10^{5}$) in the presence of atom loss. The inset shows that the atom number decreases exponentially (the blue dotted curve), which reduces the population of the upper (lower) level before (after) the superradiance pulse. The calculated pulses agree well with the experimental results (the noisy curves) with a fitted atom loss rate of $\gamma_{loss}=5.58$ Hz (compatible with the magnitude estimated in \cite{MatthewAN}) and negligible dephasing rate $d_l$.  If we increase $d_l$ (up to $\gamma_{loss}$),  the calculated pulses are reduced and shifted to earlier time and thus do not match the experimental results. If we increase $\gamma_l$ by ten times, the calculated pulses are not affected so much. These results of our analysis confirm the assessment in \cite{MatthewAN} that the system dynamics is dominated by the collective decay and the atom loss process.

The above simulation indicates that all the atoms are lost in about $0.2$ second and thus there is not time for these atoms to establish the coherence leading to the lasing with millihertz line-width. However, by feeding new atoms, we can achieve steady-state superradiance. To calculate the corresponding spectrum, we apply the quantum regression theorem and calculate the spectrum with  the Fourier transform of the two-time correlation function of the collective atomic raising and lowering operators, $S\left(\omega\right)\propto\Gamma_{c}\textrm{Re}\int_{0}^{\infty}d\tau e^{-i \omega \tau}\left\langle \sigma^{+}\left(\tau\right)\sigma^{-}\left(0\right)\right\rangle$. We follow \cite{KMoeler} to calculate the correlation function with the following procedure: we first propagate the stochastic wave-functions to long time (to approach steady-state), and then apply four different combinations of the collective lowering operator and the unit operator on the final wave-functions to initialize four ancillary wave-functions that we propagate with the MCWF method to obtain correlation function trajectories. The average of many such trajectories yields the exact correlation function. For more details, please refer to the Appendices.  Note that during the evolution the ancillary wave-functions occupy a coherent superposition of only two states (with identical $J$ and $M$ differing by unity). They can thus be propagated by solution of only two coupled equations, and their quantum jumps are also readily implemented and retain their simple two-components form at all times.

The left panel of Fig. \ref{fig:steady-state-superradiance} shows the intensity and line-width of the steady-state superradiant spectrum from $10^{5}$ strontium atoms, which are pumped individually with increasing rate $\kappa_l$, in the absence of atom loss and feeding.  We see that the intensity increases linearly with the pumping, while the linewidth decreases and approaches a minimum and then increases due to the pump-induced noise. The intensity agrees qualitatively with the prediction in \cite{DMeiser}, while the linewidth broadening occurs for much smaller pumping because of the negligible dephasing in our simulations. Note that the minimal linewidth achieved is about $16$ mHz and may be much smaller for systems with larger atom number.

The right panel of Fig. \ref{fig:steady-state-superradiance} shows the results in the presence of atom loss and feeding (in the lower state) being able to  maintain the large steady-state atom number, $N_s = 10^5$ with $\kappa_{feed} = \gamma_{loss} N_s$. As each atom typically stays for only a fraction of a second in the cavity, the incoherent pumping has to be fast to ensure excited state population available for the lasing process. Our calculations, indeed, show a threshold effect, requiring a pumping rate $\kappa_l >2$ Hz to yield a collective emission signal.   Since the adjacent Dicke states in the ancillary wave-functions are exposed to the same jumps during the correlation function evolution, their coherence survives longer than the atoms, leading to sub-hertz linewidth for pumping smaller than $9$ Hz and a minimal $60$ mHz  line-width for the $2$ Hz pumping rate. 

\paragraph*{Conclusions}
We have developed a Monte Carlo wave-function (MCWF) method to study superradiance of two-level atoms in a cavity. Our treatment incorporates the atomic collective decay, individual decay, pumping and dephasing as quantum jumps within or among Dicke ladders of fixed atom number,  and the atom loss and feeding as jumps between Dicke states with reduced and increased atom number. The wave-function populates only one Dicke state at a time and the above processes together determine how the wave-function explores the Dicke ladders. Our method is verified by the comparison with the exact master equation calculations for tens of atoms and our calculated superradiance pulses by more than $10^5$ strontium atoms agree with the experimental results with a fitted atom loss rate. Atom loss prevents steady-state lasing, but by feeding ground-state atoms into the cavity, we can stabilize the mean atom number and obtain steady-state  spectra with a minimal linewidth about $60$ mHz. Notice that the corresponding coherence time is about $16$ second and is thus hundred times longer than the time spent by any typical atom in the cavity. The MCWF unraveling of the density matrix as illustrated here can provide not only numerical results for superradiant lasing (in cross-over regime \cite{DATieri}), but may also be applied to study superradiant beats \cite{MANorcia2}, synchronization \cite{JMWeiner}, and spin-squeezing \cite{OHosten} of atomic ensembles. Furthermore, by visualizing the dynamical evolution of the quantum states, it can provide important insight into the underlying physical processes.

This work was supported by the Villum Foundation (Y. Zhang and K. M{\o}lmer) and
the European Union's Horizon 2020 research and innovation program
(No. 712721, NanoQtech, Y. X. Zhang and K. M{\o}lmer) .

\appendix

\pagebreak
\begin{center}
\textbf{\large Appendices}
\end{center}
\setcounter{equation}{0}
\setcounter{figure}{0}
\setcounter{table}{0}
\makeatletter

In the main text, we summarize briefly how Monte-Carlo wave-functions  $\left|\psi_i \left(t\right)\right\rangle$, defined by Eq. (\ref{eq:Wave-Function}) as superposition states in an effective, symmetric Dicke states $|J,M\rangle$  basis, are propagated under the influence of collective and individual dissipation. As indicated there, in the special case, where there  is no initial superposition between states of different $J$,  such superposition will never appear because of the characteristics of the quantum jumps and thus the wave-functions can always be expanded in only one single ladder of Dicke states. In the more peculiar case as studied in the main text, where the system occupies one Dicke state initially, the wave-functions will always populate one state and thus their dynamics can be effectively visualized as incoherent jumps of the numbers $J(t),M(t)$ [and the atom number $N(t)$]. In these Appendices, we offer details of the wave-functions propagation for the general case rather than the special cases to ensure that the formalism is ready to apply to other problems as indicated in the conclusion of the main text. For simplicity we will suppress the index `$i$' in the following.

\section{Monte-Carlo Wave-function Evaluation}
 The wave-functions are propagated according to the Schr{\"o}dinger equation with a non-Hermitian Hamiltonian $\overline{H}$ incorporating the atomic Hamiltonian $H_a$ and the anti-commutator part of dissipative superoperators $\mathcal{D}^{c}[\rho]$ and $\mathcal{D}^{l} [\rho]$ in the master equation. To first order in the time step $\delta t$
we have the explicit equations for the state amplitudes in the Dicke states basis
\begin{align}
&  C_{JM}\left(t+\delta t\right)=\left(1-i\tilde{\omega}_{JM}\delta t\right)C_{JM}\left(t\right). \label{eq:equation-expansion-coefficients}
\end{align}
Here the complex energy term $\tilde{\omega}_{JM}$ has the real part $\omega_{a}M+\omega_{s}\left(A_{-}^{JM}\right)^{2}$, where $\omega_a$ is the atomic transition frequency and $\omega_s$ is the cavity-induced collective Lamb shift,  and the imaginary part $-\bigl[\Gamma_{c}\left(A_{-}^{JM}\right)^{2}$$+\gamma_l \left(N/2+M\right)+\kappa_l \left(N/2-M\right)+d_l N\bigr]/2$, where, $\Gamma_c$, $\gamma_l$, $\kappa_l$, and $d_l$ denote the collective decay rate, and the individual decay, pumping and dephasing rate, respectively. The factors $A_{\pm}^{JM}=\sqrt{\left(J\mp M\right)\left(J\pm M+1\right)}$ in Eq. (\ref{eq:equation-expansion-coefficients}) depend on the Dicke state quantum numbers. These numbers govern the so-called no-jump dynamics of the wave function. The imaginary part reduces the norm of the state vector by a sum of infinitesimal probabilities $\sum_{\beta} \delta p_{\beta}$, which are, in the master equation, compensated by (sandwich) feeding terms, e.g., $\sigma^{-}\rho\left(\sigma^{-}\right)^{+}$ in $\mathcal{D}^{c}\left[\rho\right]$. In the Monte Carlo wave-function method, the feeding terms are represented by the return of the population by quantum jumps into different final states. $\beta$ enumerates eighteen different quantum jump channels, depicted in Fig. \ref{fig:system-process} (b-g) in the main text  and with the probabilities summarized in Table \ref{jumps}. The values of the different $\delta p_{\beta}$ and the corresponding quantum jump actions, will be detailed in the following.

For the collective decay, we apply the collective jump operator $\sigma^{-}$ acting on the wave-function to yield $\left|\psi\left(t+\delta t\right)\right\rangle =\sqrt{\delta t \Gamma_c} \sigma^{-}\left|\psi\left(t\right)\right\rangle $, which leads to the relation for the state amplitudes
\begin{equation}
C_{JM-1}\left(t+\delta t\right)  = \sqrt{\delta t \Gamma_c}  A_{-}^{JM}C_{JM}\left(t\right).
\label{eq:collective-decay-jump}
\end{equation}
This quantum jump, illustrated in Fig. \ref{fig:system-process} (b) in the main text, which lowers $M$ and retains $J$, occurs with the probability $\delta p_c =  \delta t \Gamma_c d_N^J \sum_M |A_{-}^{JM}C_{JM}|^2 $. Since the system has vanishing coherences between states with different $J$, we can restrict the expansion of the quantum state to a single value of $J$, which is unchanged by the collective decay process (and by the no-jump dynamics).

The decay of individual atoms lowers $M$ by unity and changes $J$ by 0 or $\pm 1$ as sketched in Fig. \ref{fig:system-process} (c) in the main text. It is represented by the master equation terms $\sum_{k}\sigma_{k}^{q}\rho\left(\sigma_{k}^{q}\right)^{+}$ ($q=z,+,-$) in $\mathcal{D}^{l}\left[\rho\right]$. Their effect on the states, $\sum_{k}\sigma_{k}^{q}\left|JM\right\rangle \left\langle JM'\right|\left(\sigma_{k}^{q}\right)^{+}$, is evaluated in \cite{BAChase,BAChase1}. Using Clebsch-Gordan (CG) expansion to express Dicke states of $N$ atoms as the product of the states of $N-1$ atoms and those of single atom with  $J\pm 1/2$ and $1/2$, respectively, one identifies for each dissipative mechanism a sum of three terms related to Dicke states with  $J_s=J+s$  ($s=0,\pm 1$). We simulate the effect of these three terms in the master equation by corresponding quantum jumps of our state vector. Note that the jumps do not correspond to a definite measurement process and outcome but are utilized here merely as a computational tool to simulate the decay of individual atoms in the master equation without embarking into the population of non-symmetric states.
In our symmetrical average evolution, the decay of individual atoms thus leads to the transformation of state amplitudes
\begin{equation}
C_{J_{s}M-1}\left(t+\delta t\right)=\sqrt{\delta t\gamma_l} P_{-,s}^{JM}C_{JM}\left(t\right),\label{eq:individual-decay-jump}
\end{equation}
implemented with the jump probabilities $\delta p_{\gamma,s}= \delta t  \gamma_l d_N^J \sum_M |P_{-,s}^{JM}C_{JM}\left(t\right)|^2 $, where the values of $P_{-,s}^{JM}$ are presented in Table \ref{jumps}. 

The incoherent excitation (pumping) of individual atoms sketched in Fig. \ref{fig:system-process} (d) in the main text leads to a similar transformation of state amplitudes
\begin{equation}
C_{J_{s}M+1}\left(t+\delta t\right)=\sqrt{\delta t\kappa_l} P_{+,s}^{JM}C_{JM}\left(t\right), \label{eq:individual-pumping-jump}
\end{equation}
implemented with the jump probabilities $\delta p_{\kappa,s}= \delta t \kappa_l d_N^J \sum_M |P_{+,s}^{JM}C_{JM}\left(t\right)|^2$, 
where the values of $P_{+,s}^{JM}$ are presented in Table \ref{jumps}.
 
The dephasing of individual atoms retains the value of $M$ and allows changes of $J$ by $0, \pm1$, as sketched in Fig. \ref{fig:system-process} (e) in the main text, This process is represented by quantum jumps with the transformation of state amplitudes
\begin{equation}
C_{J_{s}M}\left(t+\delta t\right)=\sqrt{\delta t d_l} P_{z,s}^{JM}C_{JM}\left(t\right),\label{eq:individual-pure-dephasing-jump}
\end{equation}
implemented with the jump probabilities $\delta p_{d,s}= \delta t d_l d_N^J \sum_M |P_{z,s}^{JM}C_{JM}\left(t\right)|^2 $, 
where the values of $P_{z,s}^{JM}$ are presented in Table \ref{jumps}. 

To describe the atom loss,  we use the CG expansion $|JM\rangle = \sum_{j_1,j_2}\sum_{m_1,m_2} {}^{JM} C^{j_1 m_1}_{j_2 m_2} |j_1 m_1\rangle |j_2 m_2\rangle$ with  the CG coefficients $ {}^{JM} C^{j_1 m_1}_{j_2 m_2}$ and interpret each term in the expansion as one quantum jump, which removes $2 j_1 $ atoms in the states $|j_1 m_1\rangle $ from the system and leave the $2j_2=N-2j_1$ atoms in the states $|j_2 m_2\rangle $. The different jumps will take care of all possible ways of removing the atoms in all possible states. To describe the feeding of new, uncorrelated atoms into the ensemble, we use the inverse CG expansion $ |j_1 m_1 \rangle |J M\rangle =  \sum_{j_2,m_2}  {}^{j_2 m_2} C^{j_1 m_1}_{J M}  |j_2 m_2\rangle$ and interpret each term in the expansion as one quantum jump, which adds $2 j_1$ atoms in the state $|j_1 m_1\rangle $ to the states $|J M\rangle$ of $N$ atoms to form the new states   $|j_2 m_2\rangle $ of $2j_2=N+2j_1$ atoms. The different jumps will account for all possible final states by adding atoms in one specified state. Since the probabilities of the jumps are proportional to the squared CG coefficients, the jumps of removing and feeding a single atom are more favored than other jumps. In the case of a single atom loss, i.e. $j_1 = 1/2$, we will have four different quantum jumps, which remove the single atom in the upper (lower) level and leave the $N-1$ atoms in the Dicke states of $J\pm 1/2$ and $M-1/2$ ($M+1/2$), cf.  Fig. \ref{fig:system-process} (f). These jumps can be implemented with 
\begin{equation}
C_{J_t M_s} \left(t+\delta t\right) =\sqrt{\delta t \gamma_{loss}} L_{t,s}^{JM} C_{JM} \left(t\right).\label{eq:atom-loss}
\end{equation}
Here, we have introduced $J_{t=\pm} = J \pm1/2$ and $M_{s=\pm} = M \pm1/2$, and the values of $L_{t,s}^{JM}$ are provided in Table \ref{jumps}. The jump probabilities are $\delta p_{l,ts} =  \delta t \gamma_{loss} d_N^J \sum_M |L_{t,s}^{JM} C_{JM} \left(t\right)|^2$.
In the case of single atom feeding, i.e. $j_1 = 1/2$, we will also have four different quantum jumps, which add the single atom in the upper (lower) level and leave the $N+1$ atoms in the Dicke states of $J\pm 1/2$ and $M+1/2$ ($M-1/2$),  cf.  Fig. \ref{fig:system-process} (g). These jumps can be implemented with
\begin{equation}
C_{J_t M_s} \left(t+\delta t\right) =\sqrt{\delta t \kappa_{feed}} F_{t,s}^{JM} C_{JM} \left(t\right),\label{eq:atom-feeding} 
\end{equation}
where the coefficients $F_{t,s}^{JM}$ are given in Table \ref{jumps}. The different processes occur with the jump probabilities  $\delta p_{\kappa,ts} = P_s \delta t \kappa_{feed}  d_N^J \sum_M |F_{t,s}^{JM} C_{JM} \left(t\right) |^2 $. For the atom loss we can not control the state of the lost atom and thus should consider all the possibilities. However, for the atom feeding, we can control the jumps by preparing the fed atom in the upper or lower level. For this reason, we introduce the probability $P_s$ to describe the fed atom in the upper ($s=+$) or lower ($s=-$) level, respectively.

\begin{table*}[t]
\centering{}\caption{Comprehensive list of quantum jumps associated with the different dissipative processes. The table lists the change in number of atoms and Dicke collective spin quantum numbers, the respective jump probabilities in a short time interval $\delta t$, and explicit expressions of the abbreviations used. $P_+$ and $P_-$  are the upper and lower state occupation of atoms in the feeding process. \label{jumps}}
\begin{tabular}{|c|c|c|c|}
\hline
Dissipations & Jumps & Probabilities & Abbreviations\tabularnewline
\hline
Collective Decay & $\left(N,J,M\right)\to\left(N,J,M-1\right)$ & $\delta t\Gamma_{c} d_N^J \left(A_{-}^{JM}\right)^{2}$ & $A_{-}^{JM}=\sqrt{\left(J+ M\right)\left(J- M+1\right)}$ \tabularnewline
\hline
Individual Decay & $\left(N,J,M\right)\to\left(N,J,M-1\right)$ & $\delta t\gamma_l d_N^J  \left(P_{-,0}^{JM}\right)^{2}$ & $P_{-,0}^{JM}=\sqrt{\frac{2+N}{4J\left(J+1\right)}}A_{-}^{JM}$\tabularnewline
\hline
 & $\left(N,J,M\right)\to\left(N,J-1,M-1\right)$ & $\delta t\gamma_l d_N^J \left(P_{-,-}^{JM}\right)^{2}$ & $P_{-,-}^{JM}=-\sqrt{\frac{\left(N+2J+2\right)\left(J+M\right)\left(J+M-1\right)}{4J\left(2J+1\right)}}$\tabularnewline
\hline
 & $\left(N,J,M\right)\to\left(N,J+1,M-1\right)$ & $\delta t\gamma_l d_N^J \left(P_{-,+}^{JM}\right)^{2}$ & $P_{-,+}^{JM}=\sqrt{\frac{\left(N-2J\right)\left(J-M+1\right)\left(J-M+2\right)}{4\left(J+1\right)\left(2J+1\right)}}$\tabularnewline
\hline
Individual Pumping & $\left(N,J,M\right)\to\left(N,J,M+1\right)$ & $\delta t \kappa_l d_N^J \left(P_{+,0}^{JM}\right)^{2}$ & $P_{+,0}^{JM}=\sqrt{\frac{2+N}{4J\left(J+1\right)}}A_{+}^{JM}$\tabularnewline
\hline
 & $\left(N,J,M\right)\to\left(N,J-1,M+1\right)$ & $\delta t \kappa_l d_N^J  \left(P_{+,-}^{JM}\right)^{2}$ & $P_{+,-}^{JM}=\sqrt{\frac{\left(N+2J+2\right)\left(J-M\right)\left(J-M-1\right)}{4J\left(2J+1\right)}}$\tabularnewline
\hline
 & $\left(N,J,M\right)\to\left(N,J+1,M+1\right)$ & $\delta t\kappa_l d_N^J \left(P_{+,+}^{JM}\right)^{2}$ & $P_{+,+}^{JM}=-\sqrt{\frac{\left(N-2J\right)\left(J+M+1\right)\left(J+M+2\right)}{4\left(J+1\right)\left(2J+1\right)}}$\tabularnewline
\hline
Individual Dephasing & $\left(N,J,M\right)\to\left(N,J,M\right)$ & $\delta t d_l d_N^J \left(P_{z,0}^{JM}\right)^{2}$ & $P_{z,0}^{JM} = \sqrt{\frac{2+N}{4J\left(J+1\right)}}M$\tabularnewline
\hline
 & $\left(N,J,M\right)\to\left(N,J-1,M\right)$ & $\delta t d_l d_N^J  \left(P_{z,-}^{JM}\right)^{2}$ & $P_{z,-}^{JM} = \sqrt{\frac{\left(N+2J+2\right)\left(J-M\right)\left(J+M\right)}{4J\left(2J+1\right)}}$\tabularnewline
\hline
 & $\left(N,J,M\right)\to\left(N,J+1,M\right)$ & $\delta t d_l d_N^J \left(P_{z,+}^{JM}\right)^{2}$ & $P_{z,+}^{JM}=\sqrt{\frac{\left(N-2J\right)\left(J+1-M\right)\left(J+1+M\right)}{4\left(J+1\right)\left(2J+1\right)}}$\tabularnewline
\hline
Atom Loss & $\left(N,J,M\right)\to\left(N-1,J+1/2,M+1/2\right)$ & $\delta t\gamma_{loss} d_N^J  \left(L_{+,+}^{JM}\right)^{2}$ & $L_{+,+}^{JM}=\sqrt{\frac{\left(N/2-J\right)}{N\left(2J+3\right)\left(J+M+1\right)}}$ \tabularnewline
\hline
 & $\left(N,J,M\right)\to\left(N-1,J+1/2,M-1/2\right)$ & $\delta t\gamma_{loss} d_N^J  \left(L_{+,-}^{JM}\right)^{2}$ & $L_{+,-}^{JM}=\sqrt{\frac{\left(N/2-J\right)}{N\left(2J+3\right)\left(J-M+1\right)}}$ \tabularnewline
\hline
 & $\left(N,J,M\right)\to\left(N-1,J-1/2,M+1/2\right)$ & $\delta t\gamma_{loss} d_N^J \left(L_{-,+}^{JM}\right)^{2}$ & $L_{-,+}^{JM}=\sqrt{\frac{\left(N/2+J+1\right)(J+M)}{N\left(2J+1\right)}}$\tabularnewline
\hline
 & $\left(N,J,M\right)\to\left(N-1,J-1/2,M-1/2\right)$ & $\delta t\gamma_{loss} d_N^J \left(L_{-,-}^{JM}\right)^{2}$ & $L_{-,-}^{JM}=\sqrt{\frac{\left(N/2+J+1\right)(J-M)}{N\left(2J+1\right)}}$\tabularnewline
\hline
Atom Feeding & $\left(N,J,M\right)\to\left(N+1,J+1/2,M+1/2\right)$ & $P_+\delta t \kappa_{feed} d_N^J \left(F_{+,+}^{JM}\right)^{2}$ & $F_{+,+}^{JM}=\sqrt{\frac{2\left(N+1\right)J^{2}}{\left(N/2-J+1\right)\left(2J+1\right)^{2}\left(J+1\right)\left(J-M\right)}}$\tabularnewline
\hline
 & $ \left(N,J,M\right)\to\left(N+1,J+1/2,M-1/2\right)$ & $P_- \delta t \kappa_{feed} d_N^J \left(F_{+,-}^{JM}\right)^{2}$ & $F_{+,-}^{JM}=\sqrt{\frac{2\left(N+1\right)J^{2}}{\left(N/2-J+1\right)\left(2J+1\right)^{2}\left(J+1\right)\left(J+M\right)}}$\tabularnewline
\hline
  & $\left(N,J,M\right)\to\left(N+1,J-1/2,M+1/2\right)$ & $P_+ \delta t \kappa_{feed} d_N^J \left(F_{-,+}^{JM}\right)^{2}$ & $F_{-,+}^{JM}=\sqrt{\frac{2\left(N+1\right)\left(J+1\right)\left(J+M+1\right)}{\left(N/2+J+2\right)\left(2J+1\right)^{2}}}$\tabularnewline
\hline
 & $\left(N,J,M\right)\to\left(N+1,J-1/2,M-1/2\right)$ & $P_- \delta t \kappa_{feed} d_N^J \left(F_{-,-}^{JM}\right)^{2}$ & $F_{-,-}^{JM}=\sqrt{\frac{2\left(N+1\right)\left(J+1\right)\left(J-M+1\right)}{\left(N/2+J+2\right)\left(2J+1\right)^{2}}}$\tabularnewline
\hline
\end{tabular}
\end{table*}

\section{Steady-State Spectrum Calculation}
In the main text, we indicate how to calculate the steady-state spectrum with the MCWF method. As outlined there, the spectrum can be obtained from Fourier transformation of two-time correlation function $\left\langle \sigma^{+}\left(\tau\right)\sigma^{-}\left(0\right)\right\rangle$. According to \cite{KMoeler}, this function can be calculated as $ \left(1/4\right)\sum_{k=\pm}k\bigl[\mu_{k}\overline{c_{k}\left(\tau\right)}-i\nu_{k}\overline{d_{k}\left(\tau\right)}\bigr]$, where $\overline{\cdot}$ denote the average over many trajectories of the functions
$c_{\pm}\left(\tau\right)=\left\langle \chi_{\pm}\left(\tau\right)|\sigma^{+}|\chi_{\pm}\left(\tau\right)\right\rangle $
and $d_{\pm}\left(\tau\right)=\left\langle \lambda_{\pm}\left(\tau\right)|\sigma^{+}|\lambda_{\pm}\left(\tau\right)\right\rangle $. These functions are simulated with four ancillary wave-functions $\left|\chi_{\pm}\left(\tau\right)\right\rangle $ and $\left|\lambda_{\pm}\left(\tau\right)\right\rangle $ that are specified by the initial conditions $\left|\chi_{\pm}\left(0\right)\right\rangle =\sqrt{1/\mu_{\pm}}\left(1\pm\sigma^{-}\right)\left|\psi\left(t\right)\right\rangle $,
$\left|\lambda_{\pm}\left(0\right)\right\rangle =\sqrt{1/\nu_{\pm}}\left(1\pm i\sigma^{-}\right)\left|\psi\left(t\right)\right\rangle $.
Here, $\mu_{\pm},\nu_{\pm}$ are normalization coefficients. In the Dicke states basis, these functions  are evaluated as $\sum_{J,M}d_{N}^{J}A_{\mp}^{JM} D_{JM\mp1}^{*}\left(\tau \right)D_{JM}\left(\tau \right)$, where the coefficients $D_{JM}\left( \tau \right)$ are the state amplitudes of the ancillary wave-functions. 

In the peculiar case as studied in the main text, the ancillary wave-functions will have only two non-vanishing amplitudes at any point during the stochastic wave-function evaluation. In the evaluation, these amplitudes, which reflect superposition of two adjacent Dicke states, persist despite of the quantum jumps, which ensures that the coherence time can be much longer than the time spent by any typical atom. This guarantees the lasing with millihertz line-width as shown in the main text.

\end{document}